\documentstyle[10pt,epsf,epsfig,hangcaption,xspace,amssymb,amsfonts,amsmath,amsthm,cite,
dp_delphititle,lineno]{dp_delphi}
\makeindex
\def\DpPaperGroup{EP}
\def\DpPaperRef{2003-066}
\def\DpDate{16 September 2003}
\def\DpAuthors{DELPHI Collaboration}
\def\DpSubmit{(Accepted by Phys. Lett. B)}
\def\DpTitle{{Search for single top production via FCNC at LEP at $\sqrt{s}$=189-208~GeV}}
\def\DpComment{ }
\def\DpEMail{  }
\newcommand{\gev}{{\ifmmode \mbox{Ge\kern-0.2exV}
\else Ge\kern-0.2exV\nolinebreak\fi}}
\newcommand{\mev}{{\ifmmode \mbox{Me\kern-0.2exV}
\else Me\kern-0.2exV\nolinebreak\fi}}

\begin{document}
\makeatletter
\newcount\@tempcntc
\def\@citex[#1]#2{\if@filesw\immediate\write\@auxout{\string\citation{#2}}\fi
  \@tempcnta\z@\@tempcntb\m@ne\def\@citea{}\@cite{\@for\@citeb:=#2\do
    {\@ifundefined
       {b@\@citeb}{\@citeo\@tempcntb\m@ne\@citea\def\@citea{,}{\bf ?}\@warning
       {Citation `\@citeb' on page \thepage \space undefined}}%
    {\setbox\z@\hbox{\global\@tempcntc0\csname b@\@citeb\endcsname\relax}%
     \ifnum\@tempcntc=\z@ \@citeo\@tempcntb\m@ne
       \@citea\def\@citea{,}\hbox{\csname b@\@citeb\endcsname}%
     \else
      \advance\@tempcntb\@ne
      \ifnum\@tempcntb=\@tempcntc
      \else\advance\@tempcntb\m@ne\@citeo
      \@tempcnta\@tempcntc\@tempcntb\@tempcntc\fi\fi}}\@citeo}{#1}}
\def\@citeo{\ifnum\@tempcnta>\@tempcntb\else\@citea\def\@citea{,}%
  \ifnum\@tempcnta=\@tempcntb\the\@tempcnta\else
   {\advance\@tempcnta\@ne\ifnum\@tempcnta=\@tempcntb \else \def\@citea{--}\fi
    \advance\@tempcnta\m@ne\the\@tempcnta\@citea\the\@tempcntb}\fi\fi}
 
\makeatother
\begin{titlepage}
\pagenumbering{roman}
\CERNpreprint{\DpPaperGroup}{\DpPaperRef} 
\date{{\small\DpDate}} 
\title{\DpTitle} 
\address{\DpAuthors} 
\begin{shortabs} 
\noindent
A search for single top production  (~$e^{+}e^{-} \rightarrow t\bar{c}$~) 
via Flavour Changing Neutral Currents (FCNC)
was performed using the data taken by the DELPHI detector at LEP2. 
The data analysed have been accumulated at centre-of-mass energies ranging from 189 to 
208 GeV. Limits at 95\% confidence level were obtained on the anomalous coupling 
parameters $\kappa_{\gamma}$ and $\kappa_Z$.

\end{shortabs}
\vfill
\begin{center}
\DpSubmit \ \\ 
\DpComment \ \\
\DpEMail \ \\
\end{center}
\vfill
\clearpage
\headsep 10.0pt
\addtolength{\textheight}{10mm}
\addtolength{\footskip}{-5mm}
\begingroup
%
\newcommand{\DpName}[2]{\hbox{#1$^{\ref{#2}}$},\hfill}
\newcommand{\DpNameTwo}[3]{\hbox{#1$^{\ref{#2},\ref{#3}}$},\hfill}
\newcommand{\DpNameThree}[4]{\hbox{#1$^{\ref{#2},\ref{#3},\ref{#4}}$},\hfill}
\newskip\Bigfill \Bigfill = 0pt plus 1000fill
\newcommand{\DpNameLast}[2]{\hbox{#1$^{\ref{#2}}$}\hspace{\Bigfill}}
%
\footnotesize
\noindent
\DpName{J.Abdallah}{LPNHE}
\DpName{P.Abreu}{LIP}
\DpName{W.Adam}{VIENNA}
\DpName{P.Adzic}{DEMOKRITOS}
\DpName{T.Albrecht}{KARLSRUHE}
\DpName{T.Alderweireld}{AIM}
\DpName{R.Alemany-Fernandez}{CERN}
\DpName{T.Allmendinger}{KARLSRUHE}
\DpName{P.P.Allport}{LIVERPOOL}
\DpName{U.Amaldi}{MILANO2}
% ...
\normalsize
\endgroup
\newpage
\titlefoot{Department of Physics and Astronomy, Iowa State
     University, Ames IA 50011-3160, USA
    \label{AMES}}
\titlefoot{Physics Department, Universiteit Antwerpen,
     Universiteitsplein 1, B-2610 Antwerpen, Belgium \\
     \indent~~and IIHE, ULB-VUB,
     Pleinlaan 2, B-1050 Brussels, Belgium \\
     \indent~~and Facult\'e des Sciences,
     Univ. de l'Etat Mons, Av. Maistriau 19, B-7000 Mons, Belgium
    \label{AIM}}
\titlefoot{Physics Laboratory, University of Athens, Solonos Str.
     104, GR-10680 Athens, Greece
    \label{ATHENS}}
\titlefoot{Department of Physics, University of Bergen,
     All\'egaten 55, NO-5007 Bergen, Norway
    \label{BERGEN}}
\titlefoot{Dipartimento di Fisica, Universit\`a di Bologna and INFN,
     Via Irnerio 46, IT-40126 Bologna, Italy
    \label{BOLOGNA}}
\titlefoot{Centro Brasileiro de Pesquisas F\'{\i}sicas, rua Xavier Sigaud 150,
     BR-22290 Rio de Janeiro, Brazil \\
     \indent~~and Depto. de F\'{\i}sica, Pont. Univ. Cat\'olica,
     C.P. 38071 BR-22453 Rio de Janeiro, Brazil \\
     \indent~~and Inst. de F\'{\i}sica, Univ. Estadual do Rio de Janeiro,
     rua S\~{a}o Francisco Xavier 524, Rio de Janeiro, Brazil
    \label{BRASIL}}
\titlefoot{Coll\`ege de France, Lab. de Physique Corpusculaire, IN2P3-CNRS,
     FR-75231 Paris Cedex 05, France
    \label{CDF}}
\titlefoot{CERN, CH-1211 Geneva 23, Switzerland
    \label{CERN}}
\titlefoot{Institut de Recherches Subatomiques, IN2P3 - CNRS/ULP - BP20,
     FR-67037 Strasbourg Cedex, France
    \label{CRN}}
\titlefoot{Now at DESY-Zeuthen, Platanenallee 6, D-15735 Zeuthen, Germany
    \label{DESY}}
\titlefoot{Institute of Nuclear Physics, N.C.S.R. Demokritos,
     P.O. Box 60228, GR-15310 Athens, Greece
    \label{DEMOKRITOS}}
\titlefoot{FZU, Inst. of Phys. of the C.A.S. High Energy Physics Division,
     Na Slovance 2, CZ-180 40, Praha 8, Czech Republic
    \label{FZU}}
\titlefoot{Dipartimento di Fisica, Universit\`a di Genova and INFN,
     Via Dodecaneso 33, IT-16146 Genova, Italy
    \label{GENOVA}}
\titlefoot{Institut des Sciences Nucl\'eaires, IN2P3-CNRS, Universit\'e
     de Grenoble 1, FR-38026 Grenoble Cedex, France
    \label{GRENOBLE}}
\titlefoot{Helsinki Institute of Physics, P.O. Box 64,
     FIN-00014 University of Helsinki, Finland
    \label{HELSINKI}}
\titlefoot{Joint Institute for Nuclear Research, Dubna, Head Post
     Office, P.O. Box 79, RU-101 000 Moscow, Russian Federation
    \label{JINR}}
\titlefoot{Institut f\"ur Experimentelle Kernphysik,
     Universit\"at Karlsruhe, Postfach 6980, DE-76128 Karlsruhe,
     Germany
    \label{KARLSRUHE}}
\titlefoot{Institute of Nuclear Physics,Ul. Kawiory 26a,
     PL-30055 Krakow, Poland
    \label{KRAKOW1}}
\titlefoot{Faculty of Physics and Nuclear Techniques, University of Mining
     and Metallurgy, PL-30055 Krakow, Poland
    \label{KRAKOW2}}
\titlefoot{Universit\'e de Paris-Sud, Lab. de l'Acc\'el\'erateur
     Lin\'eaire, IN2P3-CNRS, B\^{a}t. 200, FR-91405 Orsay Cedex, France
    \label{LAL}}
\titlefoot{School of Physics and Chemistry, University of Lancaster,
     Lancaster LA1 4YB, UK
    \label{LANCASTER}}
\titlefoot{LIP, IST, FCUL - Av. Elias Garcia, 14-$1^{o}$,
     PT-1000 Lisboa Codex, Portugal
    \label{LIP}}
\titlefoot{Department of Physics, University of Liverpool, P.O.
     Box 147, Liverpool L69 3BX, UK
    \label{LIVERPOOL}}
\titlefoot{Dept. of Physics and Astronomy, Kelvin Building,
     University of Glasgow, Glasgow G12 8QQ
    \label{GLASGOW}}
\titlefoot{LPNHE, IN2P3-CNRS, Univ.~Paris VI et VII, Tour 33 (RdC),
     4 place Jussieu, FR-75252 Paris Cedex 05, France
    \label{LPNHE}}
\titlefoot{Department of Physics, University of Lund,
     S\"olvegatan 14, SE-223 63 Lund, Sweden
    \label{LUND}}
\titlefoot{Universit\'e Claude Bernard de Lyon, IPNL, IN2P3-CNRS,
     FR-69622 Villeurbanne Cedex, France
    \label{LYON}}
\titlefoot{Dipartimento di Fisica, Universit\`a di Milano and INFN-MILANO,
     Via Celoria 16, IT-20133 Milan, Italy
    \label{MILANO}}
\titlefoot{Dipartimento di Fisica, Univ. di Milano-Bicocca and
     INFN-MILANO, Piazza della Scienza 2, IT-20126 Milan, Italy
    \label{MILANO2}}
\titlefoot{IPNP of MFF, Charles Univ., Areal MFF,
     V Holesovickach 2, CZ-180 00, Praha 8, Czech Republic
    \label{NC}}
\titlefoot{NIKHEF, Postbus 41882, NL-1009 DB
     Amsterdam, The Netherlands
    \label{NIKHEF}}
\titlefoot{National Technical University, Physics Department,
     Zografou Campus, GR-15773 Athens, Greece
    \label{NTU-ATHENS}}
\titlefoot{Physics Department, University of Oslo, Blindern,
     NO-0316 Oslo, Norway
    \label{OSLO}}
\titlefoot{Dpto. Fisica, Univ. Oviedo, Avda. Calvo Sotelo
     s/n, ES-33007 Oviedo, Spain
    \label{OVIEDO}}
\titlefoot{Department of Physics, University of Oxford,
     Keble Road, Oxford OX1 3RH, UK
    \label{OXFORD}}
\titlefoot{Dipartimento di Fisica, Universit\`a di Padova and
     INFN, Via Marzolo 8, IT-35131 Padua, Italy
    \label{PADOVA}}
\titlefoot{Rutherford Appleton Laboratory, Chilton, Didcot
     OX11 OQX, UK
    \label{RAL}}
\titlefoot{Dipartimento di Fisica, Universit\`a di Roma II and
     INFN, Tor Vergata, IT-00173 Rome, Italy
    \label{ROMA2}}
\titlefoot{Dipartimento di Fisica, Universit\`a di Roma III and
     INFN, Via della Vasca Navale 84, IT-00146 Rome, Italy
    \label{ROMA3}}
\titlefoot{DAPNIA/Service de Physique des Particules,
     CEA-Saclay, FR-91191 Gif-sur-Yvette Cedex, France
    \label{SACLAY}}
\titlefoot{Instituto de Fisica de Cantabria (CSIC-UC), Avda.
     los Castros s/n, ES-39006 Santander, Spain
    \label{SANTANDER}}
\titlefoot{Inst. for High Energy Physics, Serpukov
     P.O. Box 35, Protvino, (Moscow Region), Russian Federation
    \label{SERPUKHOV}}
\titlefoot{J. Stefan Institute, Jamova 39, SI-1000 Ljubljana, Slovenia
     and Laboratory for Astroparticle Physics,\\
     \indent~~Nova Gorica Polytechnic, Kostanjeviska 16a, SI-5000 Nova Gorica, Slovenia, \\
     \indent~~and Department of Physics, University of Ljubljana,
     SI-1000 Ljubljana, Slovenia
    \label{SLOVENIJA}}
\titlefoot{Fysikum, Stockholm University,
     Box 6730, SE-113 85 Stockholm, Sweden
    \label{STOCKHOLM}}
\titlefoot{Dipartimento di Fisica Sperimentale, Universit\`a di
     Torino and INFN, Via P. Giuria 1, IT-10125 Turin, Italy
    \label{TORINO}}
\titlefoot{INFN,Sezione di Torino, and Dipartimento di Fisica Teorica,
     Universit\`a di Torino, Via P. Giuria 1,\\
     \indent~~IT-10125 Turin, Italy
    \label{TORINOTH}}
\titlefoot{Dipartimento di Fisica, Universit\`a di Trieste and
     INFN, Via A. Valerio 2, IT-34127 Trieste, Italy \\
     \indent~~and Istituto di Fisica, Universit\`a di Udine,
     IT-33100 Udine, Italy
    \label{TU}}
\titlefoot{Univ. Federal do Rio de Janeiro, C.P. 68528
     Cidade Univ., Ilha do Fund\~ao
     BR-21945-970 Rio de Janeiro, Brazil
    \label{UFRJ}}
\titlefoot{Department of Radiation Sciences, University of
     Uppsala, P.O. Box 535, SE-751 21 Uppsala, Sweden
    \label{UPPSALA}}
\titlefoot{IFIC, Valencia-CSIC, and D.F.A.M.N., U. de Valencia,
     Avda. Dr. Moliner 50, ES-46100 Burjassot (Valencia), Spain
    \label{VALENCIA}}
\titlefoot{Institut f\"ur Hochenergiephysik, \"Osterr. Akad.
     d. Wissensch., Nikolsdorfergasse 18, AT-1050 Vienna, Austria
    \label{VIENNA}}
\titlefoot{Inst. Nuclear Studies and University of Warsaw, Ul.
     Hoza 69, PL-00681 Warsaw, Poland
    \label{WARSZAWA}}
\titlefoot{Fachbereich Physik, University of Wuppertal, Postfach
     100 127, DE-42097 Wuppertal, Germany 
    \label{WUPPERTAL}}
\addtolength{\textheight}{-10mm}
\addtolength{\footskip}{5mm}
\clearpage
\headsep 30.0pt
\end{titlepage}
%
\pagenumbering{arabic} 
\setcounter{footnote}{0} %
\large
\section{Introduction}

Flavour Changing Neutral Currents (FCNC) are 
highly suppressed in the Standard Model (SM) due to the Glashow-Iliopoulos-Maiani 
(GIM) mechanism~\cite{GIM}. However, small contributions appear at one-loop level 
(Br$(t\to(\gamma,g,Z) + c(u)) < 10^{-10}$) due to the 
Cabibbo-Kobayashi-Maskawa (CKM) mixing matrix~\cite{SMFCNC}. 
Many extensions of the SM, such as 
supersymmetry \cite{supers} and multi-Higgs doublet models \cite{MHiggs}, predict the presence of  
FCNC already at tree level. Some specific models \cite{Arbuzov} give rise to 
detectable FCNC amplitudes.

The most prominent signature for direct observation of FCNC processes at LEP is 
the production of a top quark together with a charm or an up quark in the process
$e^+e^- \to t \bar{c}$~\footnote{Throughout this paper
the notation $t \bar{c}$ stands for $t \bar{c}$+$t \bar{u}$ and includes
the charge conjugate contribution as well.}. The strength of the transitions $\gamma \to 
\mbox{f}\mbox{f}'$ and $Z \to \mbox{f}\mbox{f}'$ can be described in terms of the Lagrangian given in
\cite{we}:

\begin{eqnarray}
\Gamma_{\mu}^{\gamma} & = & \kappa_{\gamma} \frac{ee_q}{\Lambda}\sigma_{\mu\nu}
\left( g_1 P_l + g_2 P_r \right) q^{\nu}, \\
\Gamma_{\mu}^Z & = & \kappa_Z \frac{e}{\sin 2\Theta_W}\gamma_{\mu} 
\left( z_1 P_l + z_2 P_r\right),
\end{eqnarray}
where  $e$ is the electron charge, $e_q$
the top quark charge, $\Theta_W$ is the weak mixing angle
and $P_l$ ($P_r$) is the left (right) handed projector. The $\kappa_\gamma$ 
and $\kappa_Z$ are the anomalous couplings to the $\gamma$ and $Z$ bosons, 
respectively.
$\Lambda$ is the new physics scale. A value of 175 GeV was used for numerical 
calculations throughout 
the paper. The relative contributions of the left and right handed currents are 
determined by the 
$g_i$ and $z_i$ constants which obey the constraints:
\begin{eqnarray}
g_1^2 + g_2^2 = 1,\;\;\; z_1^2 + z_2^2 = 1.
\end{eqnarray}
In the approach which gives the most conservative limits on the couplings, 
the interference term, which depends on $g_i$ and $z_i$, 
gives a negative contribution to the cross-section of the process
$e^+e^- \to t\bar{c}$. This corresponds to the requirement \cite{we}:
\begin{eqnarray}
  g_1 z_1 + g_2 z_2 = -1.
\end{eqnarray}

The existence of anomalous top couplings to gauge bosons allows the top to
decay through $t \to c \gamma$ and  $t \to c Z$ in addition to the dominant decay
mode $t \to b W$. This effect was taken into account in the evaluation of results.
Numerical estimates of the expected number of events taking 
into account the limits
on anomalous vertices set by the CDF collaboration \cite{CDF} 
can be found in \cite{we}.

This paper is devoted to the search for FCNC processes associated to single top production
at LEP ($e^+e^- \to t\bar{c}$). Limits are set on the anomalous couplings $\kappa_\gamma$ and 
$\kappa_Z$ in the most conservative approach. The $t$ quark is expected to decay predominantly 
into $Wb$, giving distinct signatures for the leptonic and hadronic $W$ decays.
For each decay mode a dedicated analysis was developed. In the {\it semileptonic channel} 
two jets and one isolated lepton (from the $W$ leptonic decays, $W\to l\nu_l$) were 
searched for. 
In the {\it hadronic channel} four jets were required in the event (two of them from the 
$W$ hadronic decays, $W\to qq'$). A nearly background-free signature is obtained in the 
semileptonic channel, but the branching ratio is relatively low. In the hadronic 
channel, the $W$ decays give an event 
rate about two times higher, but the background conditions are less favourable.

\section{The DELPHI data and simulated samples}

The data collected with the DELPHI detector \cite{DELPHI} at $\sqrt{s} = 189-208$ GeV,
well above the $t\bar{c}$ production threshold, were used in this analysis.
The integrated luminosity used for each centre-of-mass energy bin is given in 
Table~\ref{tab:lum}.  
The data collected in the year 2000 at energies up to 208~GeV are
split into two energy bins 205~GeV and 207~GeV for centre-of-mass energies below
and above 206~GeV, respectively. 
The 189, 192, 196, 200, 202, 205 and 207~GeV energy bins correspond to 
average centre-of-mass energies of 188.6, 191.6, 195.5, 199.5, 201.6, 
204.8 and 206.6~GeV, respectively. While for the semileptonic channel the two last energy bins were 
considered separately, they were considered together in the hadronic channel.

The background process  $e^+ e^- \rightarrow Z/\gamma \rightarrow q\bar q(\gamma)$ was 
generated with
PYTHIA 6.125 \cite{pythia}. For $\mu^+\mu^-(\gamma)$ and $\tau^+\tau^-(\gamma)$,
DYMU3~\cite{dymu3} and KORALZ 4.2~\cite{koralz} were used, respectively,
while the \hbox{BHWIDE} generator~\cite{bhwide} was used for Bhabha
events.
Simulation of four-fermion final states was performed using
EXCALIBUR~\cite{excalibur} and GRC4F~\cite{grc4f}. Two-photon interactions
giving hadronic final states were generated using TWOGAM~\cite{twogam}.
Signal events were generated by a standalone simulation program interfaced with
PYTHIA 6.125 \cite{pythia} for quark hadronization. The generation of the signal
events was performed with radiative corrections included.
The SM contribution is known to be very small 
($\mbox{Br}(t \to (\gamma,g,Z) + c(u)) < 10^{-10}$ \cite{SMFCNC}) and was not
taken into account.
Both the signal and background events were passed through the
detailed simulation of the DELPHI detector and then processed
with the same reconstruction and analysis programs as the real data.

\begin{table}[hbt]
\begin{center}
\vskip 0.45 cm
\begin{tabular}{||c||c|c|c|c|c|c|c||}
\hline
\hline
$\sqrt{s}$ (GeV) & 189 & 192 & 196 & 200 & 202 & 205 & 207 \\
\hline
\hline
Luminosity (pb~${}^{-1}$) & 151.8 & 25.9 & 76.4 & 83.4 & 40.1 & 78.8 &
84.3 \\ 
\hline
\hline
\end{tabular}
\end{center}
\caption{Luminosity collected by DELPHI and used in this analysis for each centre-of-mass energy 
(see text for details).}
\label{tab:lum}
\end{table}

\section{Hadronic channel}

In the hadronic channel, the final state corresponding to the 
single top production is characterized by four jets:
a $b$ jet from the top decay, a spectator $c$ jet and  
two other jets from the $W$ hadronic decay.

In this analysis the reconstructed charged particle tracks were required to
fulfil the following criteria\footnote{The DELPHI coordinate system has the
$z$-axis aligned along the electron beam direction, the $x$-axis pointing toward
the centre of LEP and $y$-axis vertical. R is the radius in the $(x,y)$ plane. 
The polar angle $\Theta$ is measured with respect to the $z$-axis and the
azimuthal angle $\phi$ is about $z$.}:
\begin{itemize}
\item[-] momentum $p > $ 0.4 GeV/c;
\item[-] momentum error $\Delta p/p < 1$;
\item[-] $R\phi$ impact parameter $< 4$ cm;
\item[-] $z$ impact parameter $< 10$ cm.
\end{itemize}

\noindent
Tracks seen by only the central tracking devices (Vertex Detector and Inner
Detector) were rejected. Neutral clusters were required to have an energy of at
least 400 MeV. Events with the visible energy $> 100$ GeV and at least 8
charged tracks were selected for further processing.

The information of the DELPHI calorimeters and tracking 
devices was used to classify charged particles as electrons or muons according to standard DELPHI 
algorithms  \cite{DELPHI}. A well-identified lepton was designated as a ``standard" lepton. 
Whenever some ambiguity persisted the lepton was called a ``loose" lepton. To each lepton tag there 
corresponds a given detection efficiency and misidentification probability \cite{DELPHI}. Events with 
leptons with momenta above 20 GeV/$c$, identified as at least ``standard" electrons or ``loose" muons, were 
rejected.

The LUCLUS  \cite{pythia} algorithm with $d_{join} = 6.5$ GeV/c 
was then applied to 
cluster the event into jets. Events with 4, 5, or 6 jets
were selected and forced into a 4-jet topology. Each of the three most 
energetic jets must contain at least one charged particle.
The preselection was completed by requiring the event visible energy
and combined b-tag parameter~\cite{vanina} to be greater than 130~GeV and $-1.5$, respectively.
The energies and momenta of the jets were then rescaled by applying a 
constrained fit with $\mbox{NDF}=4$ imposing four-momentum 
conservation \cite{CFIT}.

The  assignment of jets to quarks is not straightforward as the kinematics of 
the event varies strongly with the energy. Near the $t\bar c$ production threshold both 
quarks are produced at rest and the subsequent top decay ($t\rightarrow Wb$) produces a 
high momentum $b$ quark. However, at higher LEP centre-of-mass energies the $c$ quark 
becomes more energetic with momentum values  up to 30 GeV/c. Four different methods 
of jet assignment were considered:

\begin{itemize}
\item[1.] the jet with highest b-tag parameter \cite{vanina} was the $b$ jet candidate 
and the least energetic jet (among the three remaining jets) was the $c$ jet candidate;
\item[2.] the most energetic jet was the $b$ jet candidate and the least energetic one was the 
$c$ jet candidate;
\item[3.] the jet with highest b-tag parameter was the $b$ jet candidate and 
two jets were assigned to the $W$ according to the probability
of the 5-C constrained fit;
\item[4.] the most energetic jet was the $b$ jet candidate and 
two jets were assigned to the $W$ according to the probability of the 5-C
constrained fit.
\end{itemize}

All the above studies were performed and the highest efficiency for the signal 
and strongest background suppression was obtained with the first method. This method 
was used in the hadronic analysis for all centre-of-mass energies.
Method (2), well suited at the kinematic threshold of single-top production,
was less efficient at the highest LEP energies because the energy of 
the $b$ jet becomes comparable to the energies of the other jets.

\begin{figure}[hp]
\begin{center}
\vspace{-0.5cm}
\mbox{\epsfig{file=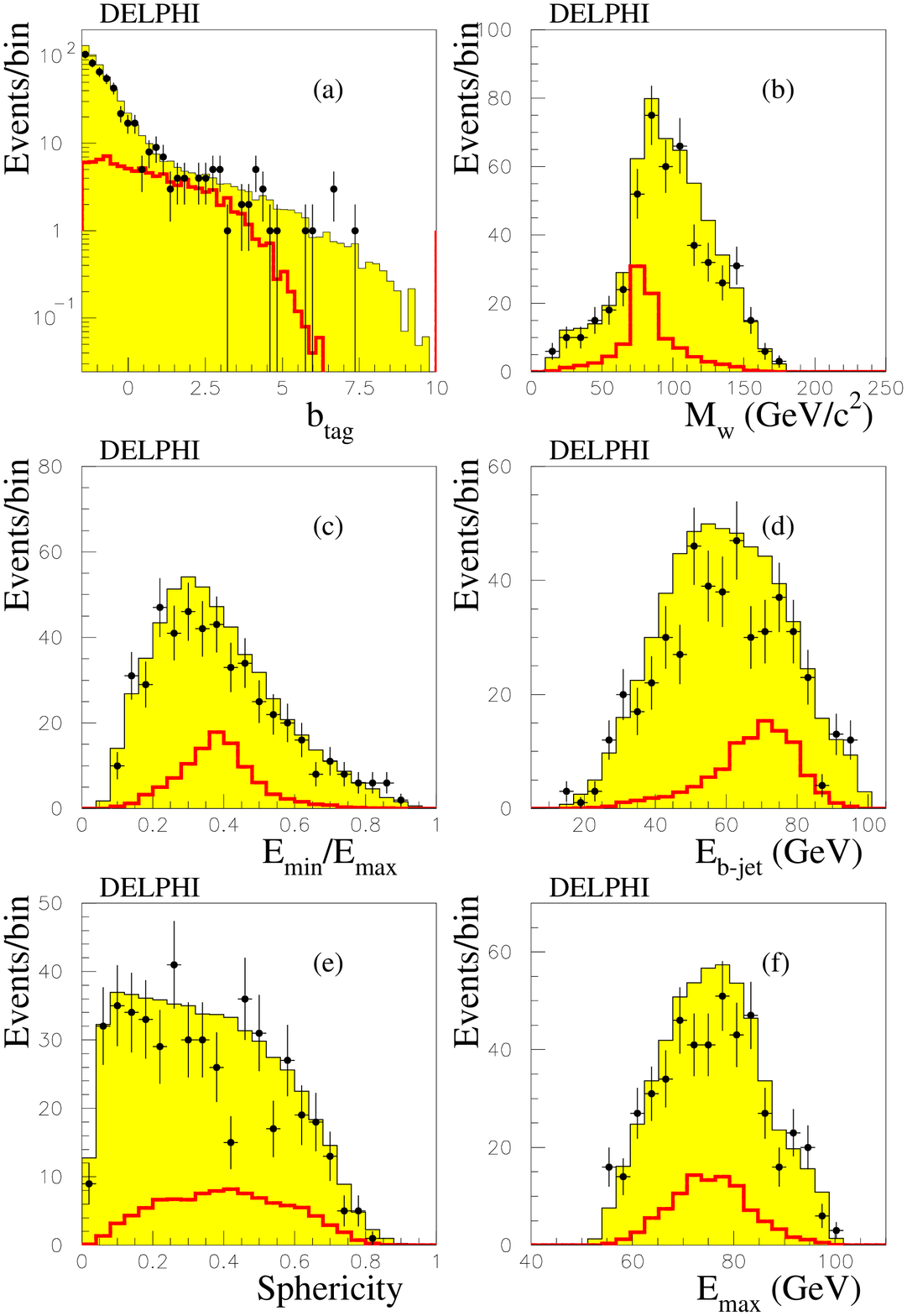,height=0.8\textheight}}
\vspace{-2mm}
\caption{Distributions of relevant variables for the hadronic decay channel after the preselection,
for $\sqrt{s}=205-207$~GeV: (a) the b-tag variable, (b)
the reconstructed $W$ mass, (c) the ratio between the minimal and the maximal jet energies,
(d) the energy of the most b-like jet, (e) the sphericity of the event and (f) the energy 
of the most energetic jet. The dots show the data, the
shaded region the SM simulation and the thick line the expected signal behaviour  
(with arbitrary normalization) for a top mass of $175$ GeV/$c^2$.}
\label{fig:oleg1}
\end{center}
\end{figure}

\begin{figure}[hp]
\begin{center}
\vspace{-1cm}
\mbox{\epsfig{file=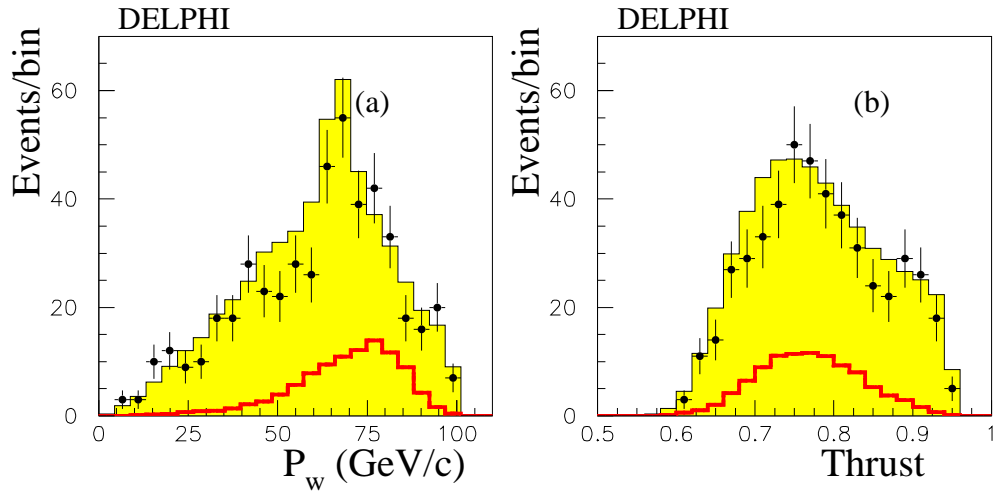,height=.85\textheight}}
\vspace{-13.cm}
\caption{Distributions of relevant variables for the hadronic decay channel after the preselection
for $\sqrt{s}=205-207$~GeV: (a) the reconstructed $W$ momentum and (b) the event thrust. The dots show the data 
and the shaded histograms show the SM simulation. The signal distribution with an arbitrary normalization  is shown 
by the thick line for a top quark mass of 175~GeV/$c^2$.}
\label{fig:oleg2}
\end{center}
\end{figure}

After the preselection, signal and background-like probabilities were assigned to each event
based on Probability Density Functions (PDF) constructed with the following variables: 


\begin{itemize}
\item the event thrust value \cite{tuning};
\item the event sphericity \cite{tuning};
\item the event b-tag calculated with the combined 
algorithm \cite{vanina};
\item the energy of the jet assigned as $b$ jet ($E_b$);
\item the energy of the most energetic jet in the event ($E_{max}$);
\item the ratio of the energies of the least  and most energetic jets
($E_{min}/E_{max}$);
\item the invariant mass of the two jets assigned as originating from the $W$
decay ($M_W$);
\item the absolute value of the reconstructed $W$ momentum ($P_W$).
\end{itemize}
Examples of these distributions are shown in figures
 \ref{fig:oleg1} and \ref{fig:oleg2}, after the preselection.

All eight PDF were estimated for the signal (${\cal P}_i^{signal}$) and
background  (${\cal P}_i^{back}$) distributions. They were used to construct the signal
${\cal{L_S}}$=$\prod_{i=1}^{8} {\cal P}_i^{signal}$ and background
${\cal{L_B}}$=$\prod_{i=1}^{8} {\cal P}_i^{back}$ likelihoods. 
A discriminant variable 
\begin{eqnarray}
W=\rm{ln}({\cal{L_S}} / {\cal{L_B}} )
\end{eqnarray}
based on the ratio of the likelihoods  was then constructed for each event.

\begin{figure}[hp]
\begin{center}
\vspace{-2cm}
\mbox{\epsfig{file=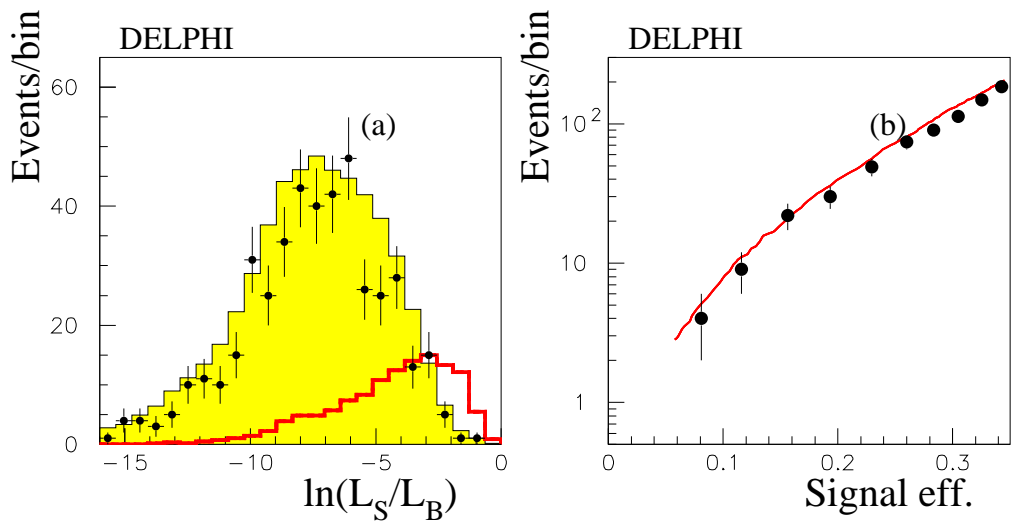,height=0.35\textheight}}
\vspace{-1.cm}
\caption{(a) distributions of the discriminant variable at $\sqrt{s}=205-207$~GeV for
data (dots), SM background simulation (shadowed region) and signal (thick line) with arbitrary
normalization and (b) number of accepted data events (dots) together with the
expected SM background simulation (full line) as a function of the signal efficiency 
(convoluted with the $W$ hadronic branching ratio) for a top mass of 175~GeV/$c^2$.}
\label{fig:oleg3}
\end{center}
\end{figure}

Figure \ref{fig:oleg3} shows 
 the discriminant variable distribution 
and  the number of accepted events, at $\sqrt{s}=205-207$~GeV, 
 as function of signal efficiency for a top mass of 
175~GeV/$c^2$. 
Events were selected by applying a cut on the discriminant variable
$\rm{ln}(\cal{L}_S$/$\cal{L}_B)$, dependent on the centre-of-mass energy. 
Its value was chosen to maximize the efficiency for a low background contamination.
The number of data events and expected background from the SM processes (mostly $WW$ background) passing
the likelihood ratio selection are shown in Table~\ref{tab:statistics2}
for all centre-of-mass energies, together with the signal efficiencies convoluted with the $W$ 
hadronic branching ratio. 
A general good agreement with the Standard Model expectations is observed.

\begin{table}[hbt]
\begin{center}
\vskip 0.45 cm
\begin{tabular}{|c||c|c||c|c|c|}
\hline
{\rm $\sqrt{s}$ } & \multicolumn{2}{c||}{Preselection} &
\multicolumn{3}{c|}{Final Selection} \\
\cline{2-6}
 (GeV)  & Data & Back.~$\pm$~stat. & Data & Back.$\pm$stat.$\pm$syst. 
& $\epsilon\times Br.$ (\%) \\
\hline
\hline
 $189$ & 568 & 530.6 $\pm$ 3.3 & 37  & 37.1 $\pm$ 1.4 $\pm$ 1.2  & 
 17.6 $\pm$ 0.5 $\pm$ 0.4 \\
\hline
 $192$ & 106 & ~91.4 $\pm$ 1.2 & 3   & ~3.4  $\pm$ 0.4 $\pm$ 0.3  & 
 17.7 $\pm$ 0.5 $\pm$ 0.4 \\
\hline
 $196$ & 266 & 253.1 $\pm$ 1.5 & 17  & 10.7 $\pm$ 0.4 $\pm$ 0.4  & 
 17.9 $\pm$ 0.6 $\pm$  0.5 \\
\hline
 $200$ & 251 & 265.0 $\pm$ 1.7 & 12  & 11.9 $\pm$ 0.5 $\pm$ 0.7  & 
 16.7 $\pm$ 0.5 $\pm$ 0.4 \\
\hline
 $202$ & 134 & 133.3 $\pm$ 0.9 & 5   & ~6.9  $\pm$ 0.3 $\pm$  0.3 & 
 17.9 $\pm$ 0.6 $\pm$ 0.5 \\
\hline
 $205-207$ & 486 & 544.1 $\pm$ 2.7 & 25  & 30.1 $\pm$ 0.9 $\pm$ 1.2  &
  17.5 $\pm$ 0.5 $\pm$ 0.6 \\
\hline
\end{tabular}
\end{center}
\caption{Number of events in the hadronic analysis at the preselection and final
selection levels, for different centre-of-mass energies. The efficiencies
convoluted with the $W$ hadronic branching ratio ($Br$) are shown for a top-quark 
mass of 175 GeV/$c^2$. Statistical and systematic errors are also given
(see section 5).}
\label{tab:statistics2}
\end{table}


\section{Semileptonic channel}

In the semileptonic channel, the final state corresponding to 
single top production is characterized by two jets (a $b$ jet from the top decay 
and a spectator $c$ jet) and at least one isolated lepton from the $W$ leptonic 
decay.

At the preselection level, events with an energy in the detector greater than 20\% of the 
centre-of-mass energy and at least 7 charged particles were selected.  The identification of muons relies on the 
association of charged particles to signals in the muon chambers and in the hadronic calorimeter and was provided 
by standard DELPHI algorithms \cite{DELPHI}.

 The identification of electrons and photons was performed by combining information from the 
electromagnetic calorimeter and the tracking system. Radiation and interaction effects were  taken 
into account by an angular clustering procedure around the main shower ~\cite{remclu}.

 Isolated leptons (photons) were defined by constructing double cones centered around the 
axis of the charged particle track (neutral cluster) with half-opening angles of 5$^\circ$ 
and 25$^\circ$ (5$^\circ$ and 15$^\circ$), and requiring that the average energy density in the region 
between the two cones was below 150~MeV/degree (100~MeV/degree), to assure isolation. In the case of 
neutral deposits, no charged particle with more than 250~MeV/c was allowed inside the
inner cone. The energy of the isolated particle was then re-evaluated as the sum of the 
energies inside the inner cone. For well identified leptons
or photons the above requirements were weakened. In this case only the external
cone was used and the angle $\alpha$ 
 was varied according to the energy of the
lepton (photon) candidate, down to 2$^\circ$ for 
$P_{lep} \ge$ 70~GeV/c (3$^\circ$
for $E_\gamma \ge$ 90~GeV), with the allowed energy inside the cone reduced 
by sin$\alpha$/sin25$^\circ$ (sin$\alpha$/sin15$^\circ$).

 Events with only one charged lepton and no isolated photons were selected. No other specific criteria 
were additionally applied to perform lepton flavour identification. 

All other particles were then forced into jets using the Durham jet algorithm~\cite{Durham}, which is 
based on a scaled transverse momentum method. Two-jet events were selected by a cut on the value of the 
corresponding resolution variable $y$ at the transition between one and two jets: $-log_{10}(y_{2 
\rightarrow 1})\ge0.45$~. The most energetic particle in each jet had to be charged. It was required 
that the momenta of the lepton and jets were greater than 10~GeV/$c$ and 5~GeV/$c$, respectively. 
Polar angles of the lepton and of the two jets were required to be in the region 20$^{\circ} \le 
\theta_{lep} \le$ 160$^{\circ}$ and 10$^{\circ} \le \theta_{j1,j2} \le$ 170$^{\circ}$, respectively. 
The missing momentum polar angle had to be above 20$^{\circ}$ and below 160$^{\circ}$ and the combined 
b-tag parameter~\cite{vanina} of the most energetic 
jet was required to be greater than $-1.1$.

 The energies and momenta of the jets, the lepton
and the momentum of the undetected neutrino (assumed to be the missing momentum) were calculated 
from four-momentum conservation with a constrained fit (NDF=1).
 Events with $\chi^2$ lower 
than 7  were accepted, provided the invariant mass of the neutrino  
and the isolated lepton was below 125~GeV/$c^2$.   
The most energetic jet was assigned to the $b$ quark 
and the second jet to the $c$ quark.
The top mass was reconstructed as the invariant mass of 
the $b$ jet, the isolated lepton and the neutrino four-momenta.

\begin{figure}[hp]
\begin{center}
\vspace{-1cm}
\mbox{\epsfig{file=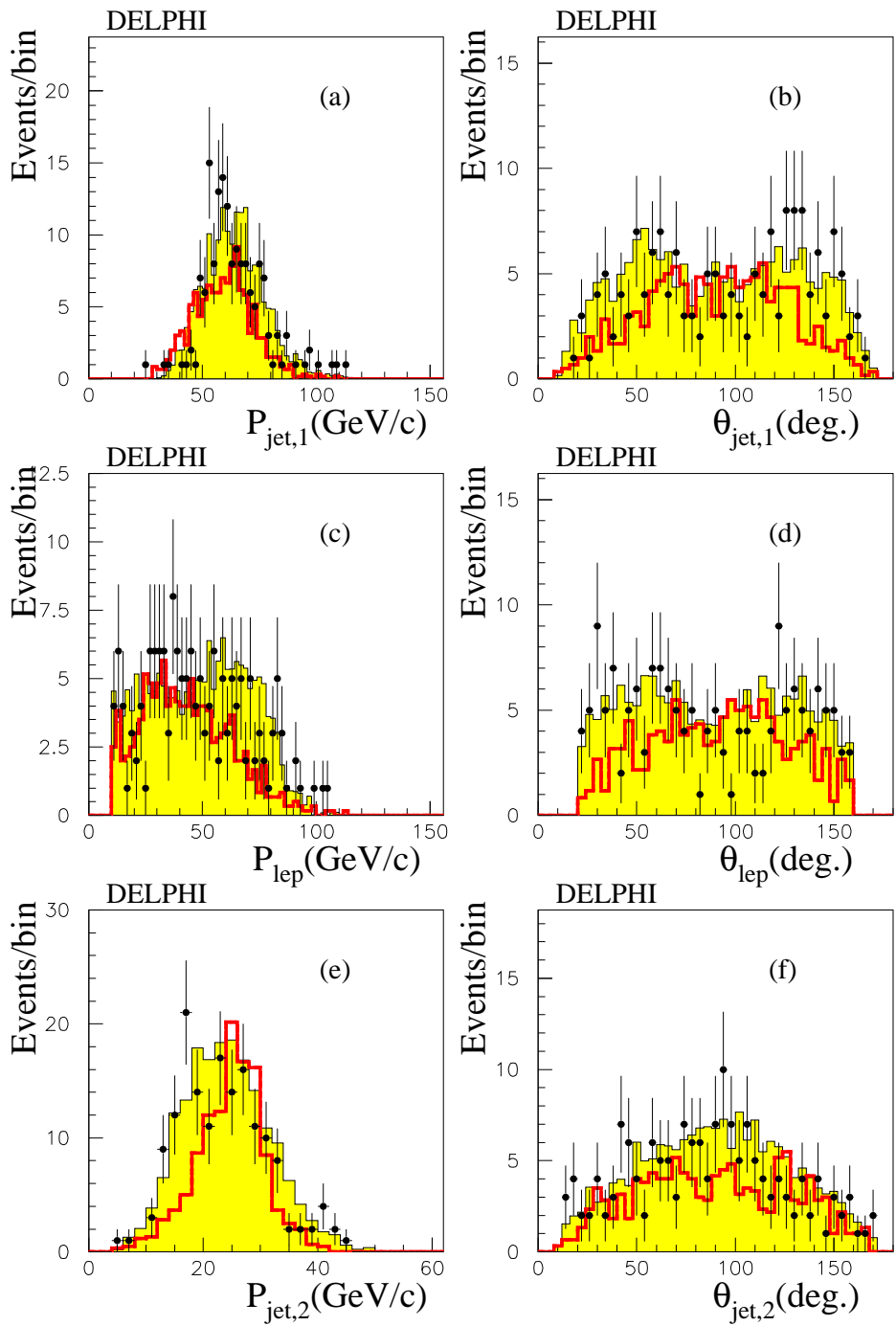,height=0.8\textheight}}
\vspace{-2mm}
\caption{Distributions of relevant variables for the semileptonic decay channel at the preselection level
for $\sqrt{s}=205-207$~GeV. 
The momentum of the most energetic jet (a) and its polar angle (b),
the lepton momentum (c) and its polar angle (d),
the momentum of the least energetic jet (e) and its polar angle (f)
are shown. The dots show the data, the
shaded region the SM simulation and the thick line the expected signal behaviour  
(with arbitrary normalization) for a top mass of $175$ GeV/$c^2$.}
\label{fig:fcnc_m15_exdis205_lev1_lslb-999._1}
\end{center}
\end{figure}

\begin{figure}[hp]
\begin{center}
\vspace{-1cm}
\mbox{\epsfig{file=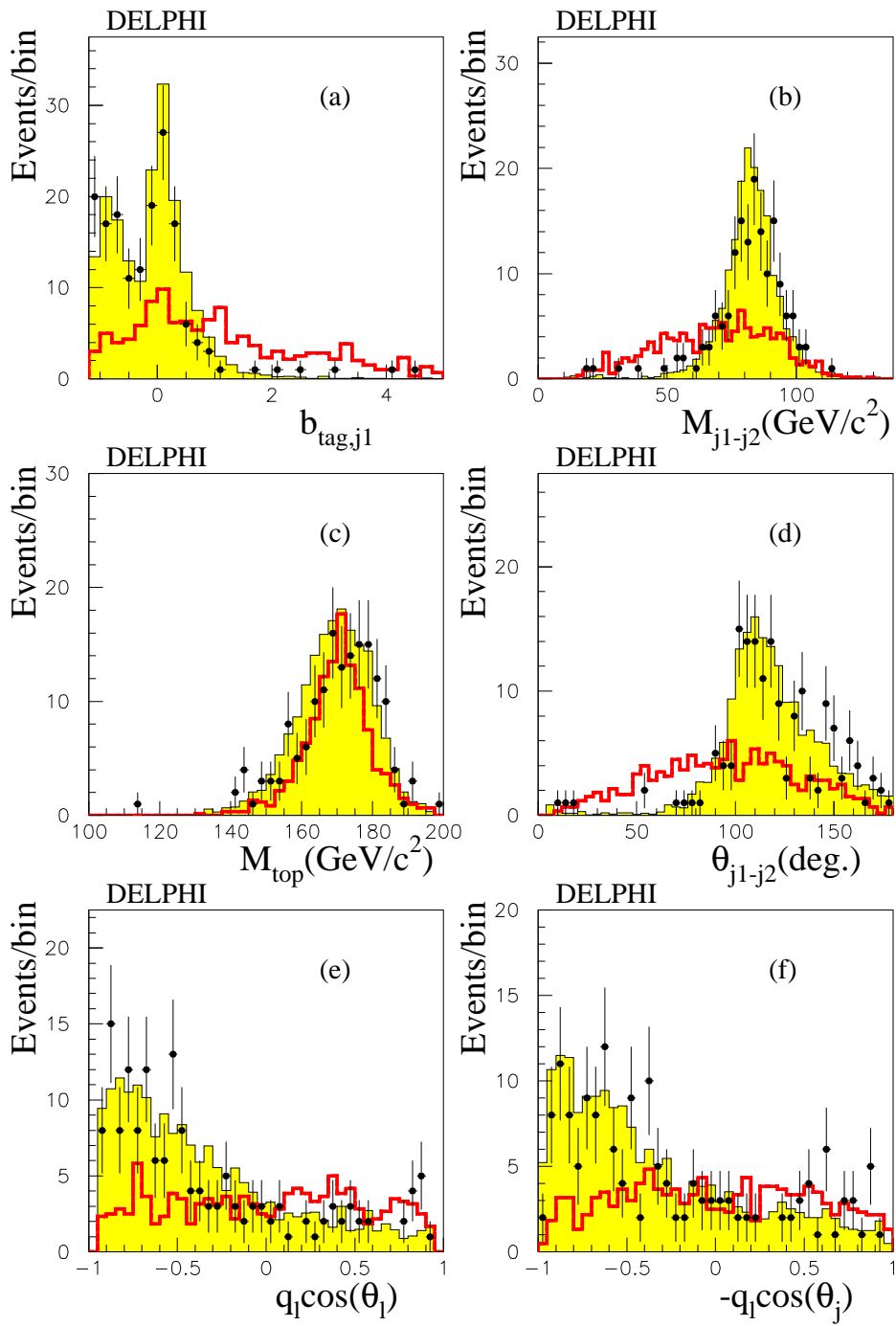,height=0.8\textheight}}
\vspace{-2mm}
\caption{Distributions of relevant variables at the preselection level in the
semileptonic decay channel, for $\sqrt{s}= 205-207$~GeV:
(a) the most energetic jet b-tag parameter, (b) the reconstructed two jet system
mass, (c) top mass, (d) the angle between the jets, (e) $\rm q_l cos(\theta _l)$  
(see text for explanation) and (f) $\rm -q_l cos(\theta _j)$. The dots show the data,
the shaded region the SM simulation and the thick line the expected signal
behaviour (with arbitrary normalization) for a top quark mass of $175$ GeV/$c^2$.}
\label{fig:fcnc_m15_exdis205_lev1_lslb-999._2}
\end{center}
\end{figure}

Figures \ref{fig:fcnc_m15_exdis205_lev1_lslb-999._1} and 
\ref{fig:fcnc_m15_exdis205_lev1_lslb-999._2} 
show some relevant distributions for data and MC, after the preselection
and for $\sqrt{s}=205-207$~GeV. The number of events at preselection and final 
selection levels are given in Table~\ref{tab:stat_semileptonic} for each 
centre-of-mass energy. Most of the background comes from SM $e^+e^-\rightarrow WW$ events.

 After the preselection, signal and background-like probabilities 
were assigned to each event (as for the hadronic channel) based on PDF constructed with the 
following variables: 

\begin{itemize}
\item momentum of the less energetic jet;
\item more energetic jet b-tag variable ~\cite{vanina};
\item reconstructed mass of the two jets;
\item reconstructed top mass;
\item angle between the two jets;
\item lepton-neutrino invariant mass;
\item $q_l \cdot \cos \theta_l$, where $q_l$ is the charge and $\theta_l$ is the
polar angle of the lepton;
\item $q_{j1} \cdot \cos \theta_{j1}$, where $q_{j1}=-q_{l}$ and $\theta_{j1}$
is the polar angle   of the more energetic jet;
\item $p_{j1} \cdot [\sqrt{s} - p_{j1}(1-\cos \theta_{j1\, j2})]$, where
$p_{j1}$ is the momentum of the
more energetic jet and $\theta_{j1\, j2}$ is the angle between the two jets. This variable
is proportional to $(m^2_t-m^2_W)/2.$ i.e., not dependent on the centre-of-mass energy.
\end{itemize}

\begin{figure}[hp]
\begin{center}
\vspace{-1cm}
\mbox{\epsfig{file=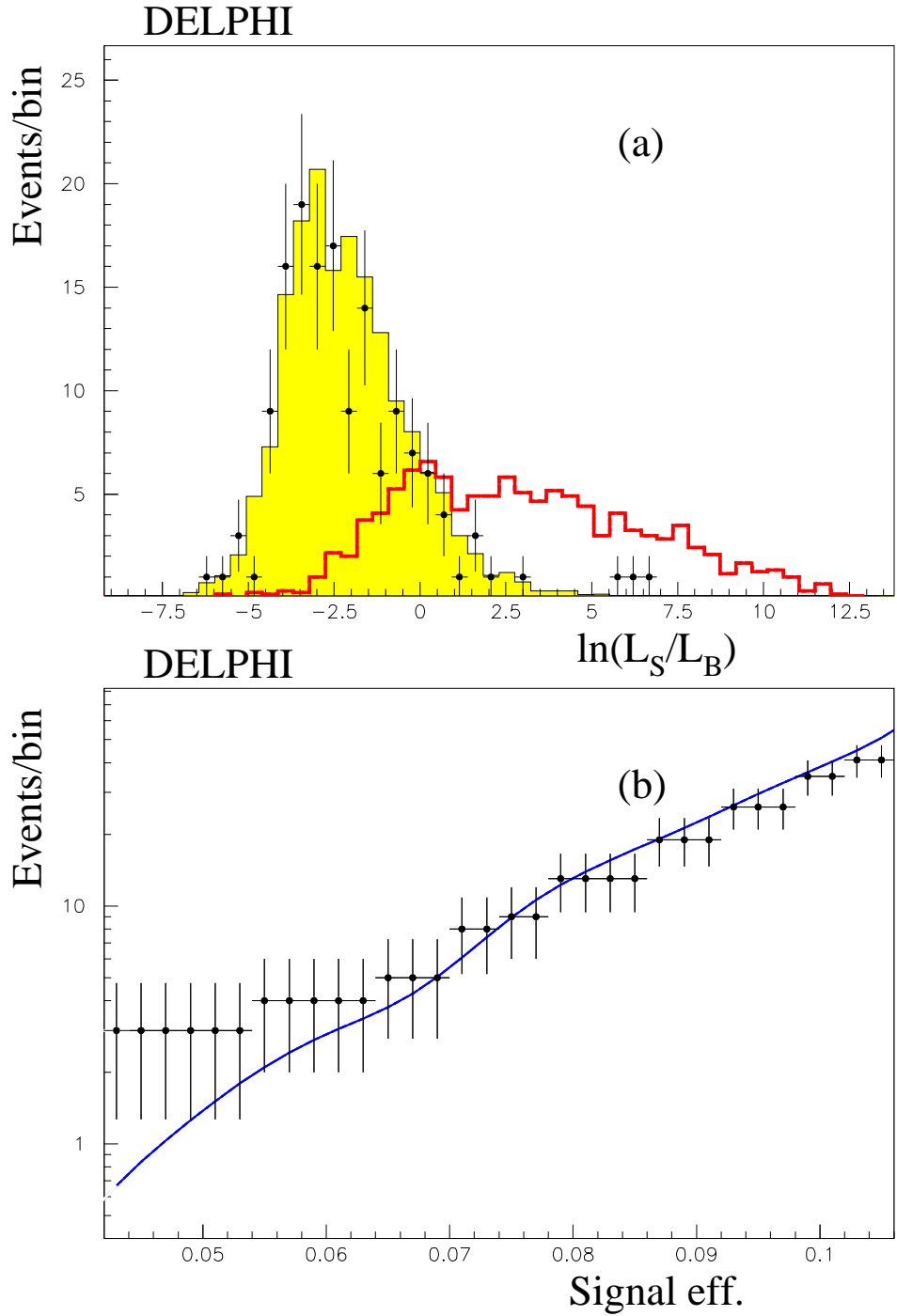,height=0.8\textheight}}
\vspace{-2mm}
\caption{(a) the discriminant variable distribution for $\sqrt{s} =205-207$~GeV is shown. The dots
show the data, the shaded region the SM simulation and the thick line the
expected signal behaviour (with arbitrary normalization) for a top quark mass of 
$175$ GeV/$c^2$. (b) number of accepted data events (dots) together with the
expected SM background simulation (full line) as a function of the signal efficiency 
(convoluted with the $W$ leptonic branching ratio) for a top mass
of 175~GeV/$c^2$.}
\label{fig:fcnc_m15_exdis205_lev1_lslb-999._3}
\end{center}
\end{figure}

The signal ($\cal{L}_S$) and background ($\cal{L}_B$) likelihoods
were used on an event-by-event basis to compute a discriminant variable defined as 
~ln$(\cal{L}_S$/$\cal{L}_B)$. 
A loose cut on the signal likelihood was applied to 
the events. 
Figure \ref{fig:fcnc_m15_exdis205_lev1_lslb-999._3} presents, after this cut,
the discriminant variable distribution and the number of events accepted as a 
function of signal efficiency for $\sqrt{s}$=205-207~GeV
(assuming a top mass of 175~GeV/$c^2$ for the signal). 
There is a general good agreement between the data and
the SM predictions. The background distribution has a tail for higher values of 
the discriminant variable which goes below every data event.
Correlations between the variables were studied. Their effect on the 
likelihood ratio is small.

Events were further selected by applying a cut on the discriminant variable ~ln$(\cal{L}_S$/$\cal{L}_B)$,
dependent on the centre-of-mass energy. Table \ref{tab:stat_semileptonic} shows the number of
data and background events which passed the cut for the different centre-of-mass energies. The
efficiencies convoluted with the $W$ leptonic branching ratio are also shown.
The dominant backgrounds come from SM $e^+e^-\rightarrow WW$ and $e^+e^-\rightarrow ~q\bar q$ events.

\begin{table}[hbt]
\begin{center}
\vskip 0.45 cm
\begin{tabular}{|c||c|c||c|c|c|}
\hline
{\rm $\sqrt{s}$ } & \multicolumn{2}{c||}{Preselection} &
\multicolumn{3}{c|}{Final Selection} \\
\cline{2-6}
 (GeV)  & Data & Back.~$\pm$~stat. & Data & Back.$\pm$stat.$\pm$syst. 
& $\epsilon\times Br.$ (\%) \\
\hline
\hline
 $189$ & 102 & 120.7 $\pm$ 4.3  & 1  & ~2.4 $\pm$ 0.7 $\pm$ 0.8  & 8.0 $\pm$ 0.3 $\pm$ 0.5 \\
\hline
 $192$ & ~24 & ~21.5 $\pm$ 0.8  & 1  & ~0.5 $\pm$ 0.1 $\pm$ 0.1  & 7.7 $\pm$ 0.9 $\pm$ 0.5 \\
\hline
 $196$ & ~72 & ~76.2 $\pm$ 2.5  & 2  & ~0.9 $\pm$ 0.3 $\pm$ 0.1  & 7.1 $\pm$ 0.9 $\pm$ 0.5 \\
\hline
 $200$ & ~95 & ~87.6 $\pm$ 2.8  & 1  & ~2.0 $\pm$ 0.5 $\pm$ 0.3  & 6.9 $\pm$ 0.3 $\pm$ 0.3 \\
\hline
 $202$ & ~40 & ~42.2 $\pm$ 1.3  & 1  & ~1.7 $\pm$ 0.3 $\pm$ 0.1  & 7.9 $\pm$ 0.4 $\pm$ 0.3 \\
\hline
 $205$ & ~90 & ~90.0 $\pm$ 2.9  & 2  & ~1.4 $\pm$ 0.4 $\pm$ 0.1  & 6.2 $\pm$ 0.3 $\pm$ 0.3 \\
\hline
 $207$ & ~71 & ~90.2 $\pm$ 2.6  & 2  & ~1.9 $\pm$ 0.5 $\pm$ 0.2  & 6.2 $\pm$ 0.3 $\pm$ 0.4 \\
\hline
\end{tabular}
\end{center}
\caption{Number of events in the semileptonic analysis at
the preselection and final selection levels, for 
the different centre-of-mass energies. The efficiencies
convoluted with the $W$ leptonic branching ratio are also
shown for a top mass of 175~GeV/$c^2$. Statistical and systematic errors
are given (see the systematic errors and limit derivation section).}
\label{tab:stat_semileptonic}
\end{table}


\section{Systematic errors and limit derivation}

Studies of systematic errors were performed and their effect evaluated at the 
final selection level. The stability of the results with respect to variations on
the selection criteria, the PDF definition, the different hadronization 
schemes
and the uncertainty in top quark mass were studied. 

Concerning the stability of the results, an independent (and large, compared to the 
resolution) variation on the selection criteria applied to analysis variables like the missing momentum 
polar angle, the combined b-tag of the 
most energetic jet, the $W$ mass, the Durham resolution 
variable, etc. was allowed.  The most significant contributions gave
a maximum error of 0.5 events and 0.3\% for the expected background and efficiency, respectively. Different
smoothing procedures were performed for the PDF definition and their effect is at most
0.5 events (0.4\%) for the expected background (signal efficiency). 
 Different hadronization schemes (string and independent)  \cite{pythia} were studied for the 
signal and their effect contributes at most 0.1\% for the signal efficiency error. 
 The uncertainty on the top quark mass is the most important source of systematic errors. It affects not
only the total production cross-section but also the kinematics of signal events. In terms of signal 
efficiency, its effect could be as high as 0.9\%  for the semileptonic channel  (in the 
mass range between 170~GeV/$c^2$ and 180~GeV/$c^2$). The effects of such variations 
(added quadratically) on the final selection criteria are quoted as a systematic error in 
Tables~\ref{tab:statistics2} and ~\ref{tab:stat_semileptonic}.

The number of data and expected SM background events for the hadronic and 
semileptonic channels, the 
respective signal efficiencies and data luminosity collected at the various centre-of-mass energies were 
combined to derive limits in the ($\kappa_{\gamma}$, $\kappa_Z$) plane using a Bayesian approach 
\cite{Obraztsov_bayes}. In total, 13 independent channels (6 in the hadronic and
7 in the semileptonic modes) correspond to different $\sqrt{s}$ values. These
channels are fitted simultaneously to extract the limits on the FCNC parameters. 
The total production cross-section and top FCNC decay widths dependence with 
$\kappa_{\gamma}$ and $\kappa_Z$ were properly considered ~\cite{we} in the limit derivation.

The effect of systematic errors on the ($\kappa_\gamma,\kappa_Z$) plane 
limits was considered. Initial State Radiation (ISR) and QCD corrections \cite{QCDcorr} were also taken into 
account in the $t\bar c$ total production cross-section. 

\begin{figure}[ht]
\begin{center}
\includegraphics[width=14.5cm]{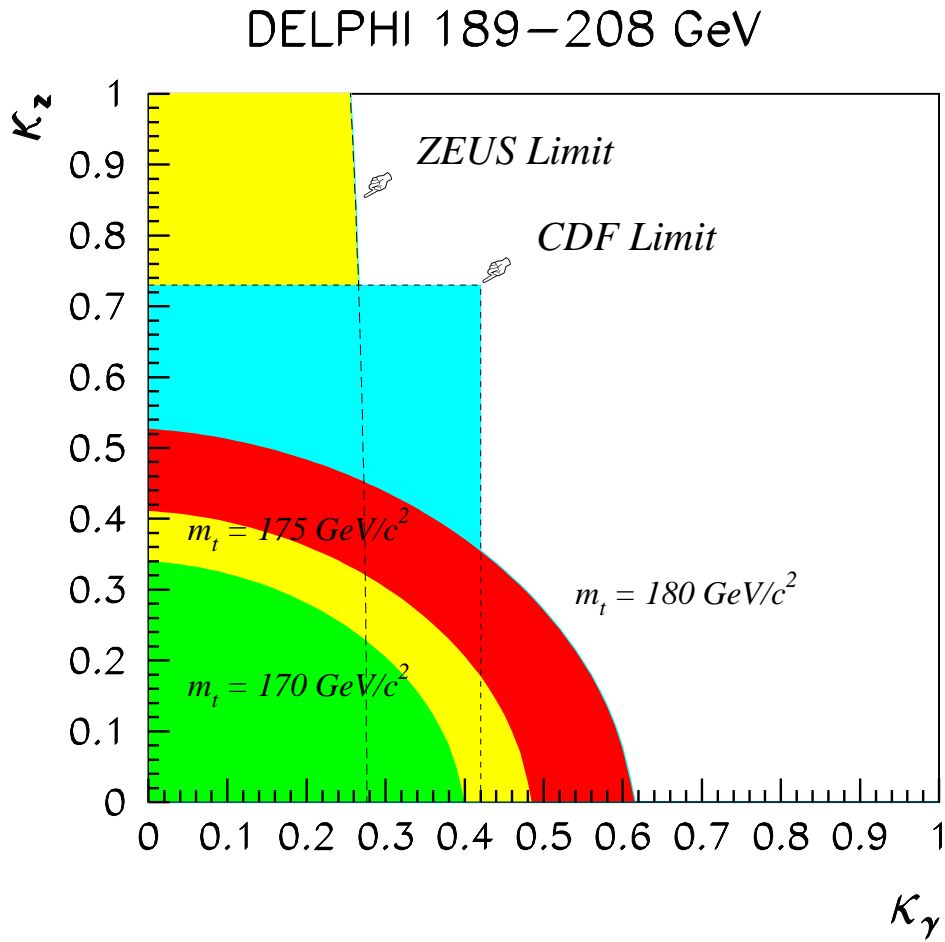}
\end{center}

\caption{Limits at 95\%\ confidence level in the $\kappa_{\gamma}-\kappa_Z$ plane. The 
different curved and filled areas represent the regions allowed by DELPHI for 
different top quark masses. Radiative corrections were taken into account in the total
production cross-section at LEP. The CDF and ZEUS allowed regions are also shown for a top quark mass of
$175$ GeV/$c^2$. The ZEUS limits are scaled by a factor of $\sqrt{2}$ because
of the difference in the Lagrangian definitions.}
\label{fig:oleg4}
\end{figure}

Figure \ref{fig:oleg4} shows the 95\% confidence level (C.L.) upper limits in the
($\kappa_{\gamma}$, $\kappa_Z$) plane obtained by this analysis. The different filled 
areas correspond to the allowed regions obtained for different top mass values and
$\Lambda=175$ GeV. Due to the $s$-channel $Z$ dominance, the LEP2 data are less sensitive to 
the  $\kappa_{\gamma}$ parameter than to $\kappa_Z$. The upper limits obtained by the CDF 
collaboration~\cite{CDF} and ZEUS \cite{others_HERA} are also shown in the figure for comparison.
The 95\%\ C.L. upper limits on each coupling parameter, setting the other coupling to 
zero, are summarized in Table \ref{tab:resu}. For comparison the values at $m_t=175$~GeV/$c^2$ are
$\kappa_Z (\kappa_{\gamma} = 0)=0.434$ and $\kappa_{\gamma} (\kappa_Z = 0)=0.505$ 
if the Born level cross-section (without radiative corrections) is taken into account.

\begin{table}[hbt]
\begin{center}
\vskip 0.45 cm
\begin{tabular}{|c|ccc|} \hline\hline 
$m_t$ (GeV/$c^2$) & 170 & 175 & 180 \\ \hline
$\kappa_Z (\kappa_{\gamma} = 0)$  &  0.340  & 0.411   & 0.527   \\
$\kappa_{\gamma} (\kappa_Z = 0)$  &  0.402  & 0.486   & 0.614   \\ \hline 
\hline
 \end{tabular}
\end{center}
\caption{95\%\ C.L. upper limits derived from the combined hadronic and semileptonic
 channels at $\sqrt{s}= 189-208$ GeV for $\Lambda=175$ GeV.}
\label{tab:resu}
\end{table}

Upper limits were also obtained by using only the hadronic and the semileptonic
channels separately when radiative corrections to the total
production cross-section were taken into account. The values at $m_t=175$~GeV/$c^2$ are
$\kappa_Z (\kappa_{\gamma} = 0)=0.491$  ($0.547$) and $\kappa_{\gamma} (\kappa_Z = 0)=0.568$ ($0.625$)
for the hadronic (semileptonic) channel alone.

\section{Summary}

The data collected by the DELPHI detector at centre-of-mass 
energies ranging from 189 to 208~GeV were used to perform a search for 
FCNC $t\bar{c}$ production, in the hadronic and semileptonic topologies.
No deviation with respect to the SM expectations was found.
Upper limits on the anomalous couplings $\kappa_{\gamma}$ 
and $\kappa_Z$ were derived.
A comparison with CDF ~\cite{CDF} and ZEUS \cite{others_HERA} is also shown. 
Results on the search for single-top production were 
also obtained by the other experiments at LEP \cite{others_LEP}.

\subsection*{Acknowledgements}
\vskip 3 mm
 We are greatly indebted to our technical 
collaborators, to the members of the CERN-SL Division for the excellent 
performance of the LEP collider, and to the funding agencies for their
support in building and operating the DELPHI detector.\\
We acknowledge in particular the support of \\
Austrian Federal Ministry of Education, Science and Culture,
GZ 616.364/2-III/2a/98, \\
FNRS--FWO, Flanders Institute to encourage scientific and technological 
research in the industry (IWT), Federal Office for Scientific, Technical
and Cultural affairs (OSTC), Belgium,  \\
FINEP, CNPq, CAPES, FUJB and FAPERJ, Brazil, \\
Czech Ministry of Industry and Trade, GA CR 202/99/1362,\\
Commission of the European Communities (DG XII), \\
Direction des Sciences de la Mati$\grave{\mbox{\rm e}}$re, CEA, France, \\
Bundesministerium f$\ddot{\mbox{\rm u}}$r Bildung, Wissenschaft, Forschung 
und Technologie, Germany,\\
General Secretariat for Research and Technology, Greece, \\
National Science Foundation (NWO) and Foundation for Research on Matter (FOM),
The Netherlands, \\
Norwegian Research Council,  \\
State Committee for Scientific Research, Poland, SPUB-M/CERN/PO3/DZ296/2000,
SPUB-M/CERN/PO3/DZ297/2000 and 2P03B 104 19 and 2P03B 69 23(2002-2004)\\
JNICT--Junta Nacional de Investiga\c{c}\~{a}o Cient\'{\i}fica 
e Tecnol$\acute{\mbox{\rm o}}$gica, Portugal, \\
Vedecka grantova agentura MS SR, Slovakia, Nr. 95/5195/134, \\
Ministry of Science and Technology of the Republic of Slovenia, \\
CICYT, Spain, AEN99-0950 and AEN99-0761,  \\
The Swedish Natural Science Research Council,      \\
Particle Physics and Astronomy Research Council, UK, \\
Department of Energy, USA, DE-FG02-01ER41155, \\
EEC RTN contract HPRN-CT-00292-2002. \\



\newpage

\end{document}